# Joule Expansion Imaging Techniques on Microlectronic Devices

S. Grauby, L-D. Patino Lopez, A. Salhi, E. Puyoo, J-M. Rampnoux, W. Claeys, S. Dilhaire

Centre de Physique Moléculaire Optique et Hertzienne,Université Bordeaux 1, 351, cours de la Libération,
33405 Talence cedex, France,
s.grauby@cpmoh.u-bordeaux1.fr

*Abstract-* We have studied the electrically induced off-plane surface displacement on two microelectronic devices using Scanning Joule Expansion Microscopy (SJEM). We present the experimental method and surface displacement results. We show that they can be successfully compared with surface displacement images obtained using an optical interferometry method. We also present thermal images using Scanning Thermal Microscopy (SThM) technique to underline that SJEM is more adapted to higher frequency measurements, which should improve the spatial resolution.

## I. INTRODUCTION

As integration density of microelectronic circuits goes increasing, there is a need for methods able to measure local temperature variations or surface displacements at submicronic scales.

As a consequence, well-known temperature measurement methods such as infrared imaging [1], liquid crystals measurements or temperature measurements using micrometric thermocouples deposited on the surface of the device[2] are not adapted to this kind of samples as they offer a bad spatial resolution (5 to 10 µm minimum) regarding the device dimensions. Moreover, a thermocouple implies a contact with the sample that can damage it or disrupt its functioning. Among the optical methods for submicronic thermal mapping, thermoreflectance[3-8] is a useful non contact and non invasive method which presents a good spatial resolution as limited by diffraction to the order of magnitude of the illuminating wavelength. Nevertheless, when studying structures as thin as a few hundreds nanometers, only a Scanning Thermal Microscope (STHM) [9-12] can theoretically reach temperature variations measurements at this scale.

As for surface displacement measurements, for the same reasons as for temperature measurements techniques, optical interferometry [8, 13, 14] is a very useful technique. In this paper, we propose to use a Scanning Joule Expansion Microscopy (SJEM) technique[15] based on an Atomic Force microscope to detect the surface displacement induced by the supplying current. We compare the results with interferometric surface displacement measurements and with thermal variations obtained by SThM.

## II. EXPERIMENTAL SET-UP AND DEVICE UNDER TEST

The SJEM technique is based on the use of a classical Atomic Force Microscope (AFM) equipped with a topographical tip. This set-up enables to measure and extract both topographical and deflection images.

The principle of SJEM (Fig. 1) is to supply the device under test with an AC sine electrical voltage which creates a modulated surface thermal expansion due to the Joule effect. The AFM system enables the control of the tip position and the contact force with the device under test (DUT). Monitoring is performed with a feedback loop between the signal of three x-y-z piezo-electrical ceramics adjusting the tip cantilever position and four photodiodes tracking a laser beam which is reflected on the mirror of the probe. The AFM photodiodes detects the cantilever deflection due to both expansion and sample topography. Since the feedback controller of the AFM has a bandwidth of about $f_c$=10 kHz, the photodiode signal below the $f_c$ cut-off frequency is processed for feedback control of the z-piezo to image surface topography. Then, one possibility [15] is to keep the Joule heating frequency above the cut-off frequency to avoid feedback response. Then, we can analyze the deflection image to extract the expansion image.

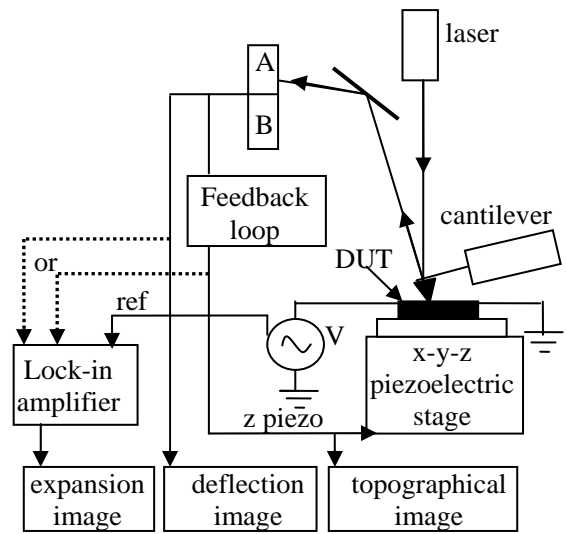

Fig. 1: SJEM set-up

Another possibility, below the cut-off frequency, is to deduce the expansion image from the topographical image. Because, below the cut-off frequency, the feedback loop adjusts the tip position and the deflection signal is equal to zero. Therefore, we cannot use the deflection image to obtain the expansion one.

Thus, depending on the Joule heating frequency, we will use either the topographical image (below $f_c$) or the deflection image (above $f_c$) as input signal for the lock-in amplifier. At last, the lock-in amplifier is tuned to the Joule heating frequency which detects only the expansion signal and provides this to an auxiliary AFM channel to form the expansion image. We can either measure the amplitude or the phase of the expansion signal. In this paper, we focus on amplitude images.

We have studied two kinds of structures (Fig. 2) constituted by a series of 9 parallel strip resistors which serve themselves as a heat source. Indeed, supplying them with a sine voltage, they are submitted to a modulated variation of temperature which creates a modulated expansion of the device. The width of each resistor is 0.35 µm. They are deposited on a silicon substrate and covered by a passivation layer made of 200nm silicon oxide which is transparent for visible wavelengths. The difference between both structures is the spacing between the resistors. In the first one named sample A (Fig. 2(a)), the distance between two consecutive resistors is 10 µm whereas it is only 0.8 µm in the second one named sample B (Fig. 2(b)). These structures are part of a whole test device designed for the evaluation of various temperature measurement techniques. Fig. 2(a) and 2(b) correspond to a 100×100 µm² surface. The die was implemented using 0.35 µm CMOS technology. The value of each of the 9 resistors is r=1845 Ω for sample A and r=2934 Ω for sample B.

### III. SCANNING JOULE EFFECT DISPLACEMENT MEASUREMENTS

These samples have been studied using SJEM technique to deduce off-plane surface displacement images. The device is submitted to a f frequency positive sine wave voltage:

$$V = V_0 (1 + \cos(2\pi f t)) \qquad (1)$$

with $V_0$ the mean amplitude voltage.

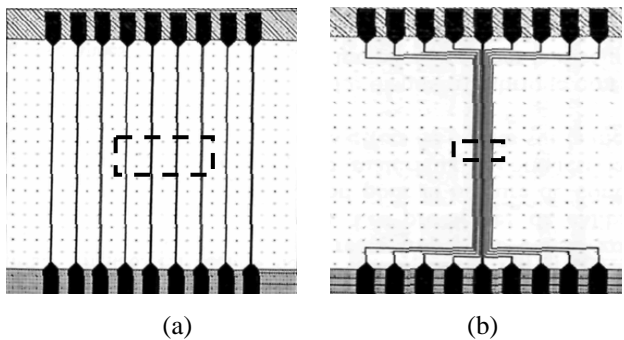

Fig. 2: Samples constituted of nine 0.35 µm thin resistors with a distance between 2 resistors of: (a) 10 µm for sample A, (b) 0.8 µm for sample B.

The power P dissipated by Joule effect in each resistor is then given by:

$$P = \frac{V^2}{r} = \frac{V_0^2}{r}\left(\frac{3}{2} + 2\cos(2\pi f t) + \frac{\cos(2\pi 2 f t)}{2}\right) \qquad (2)$$

Then, each resistor temperature variation ΔT is proportional to the dissipated power P and the local expansion signal measured on each resistor by SJEM is itself proportional to ΔT. Consequently, each frequency component of the power induces an expansion signal on the sample. Using the lock-in amplifier locked on the f frequency, we detect the f frequency expansion amplitude.

#### A. SJEM on Sample A

We present in the upper part of Fig. 3 the expansion image obtained on sample A on a 40 µm×10 µm area (dashed area in Fig. 2(a)) corresponding to 64 lines of 256 points each with $V_0$=2.5 V and f=10 kHz. Because of the frequency very close to the feedback loop cut-off one, we have deduced it from the deflection one. Under the image, Fig. 3 also presents a section perpendicularly to the resistors or more precisely the mean signal over the 64 lines of the image above. The expansion signal in volts has been translated in pm thanks to the deflection sensitivity estimated to 35 nm/V. The mean distance between 2 consecutive maxima is 10.6 µm, in good agreement with the 10.35 µm expected distance. The displacement is made of two contributions: a global "background" displacement on which a local displacement on each resistor is superimposed. The maximal displacement on the resistors reaches 360 pm for a global displacement of 320 pm. Scanning a larger area (80 µm instead of 40 µm), we can notice that the global displacement shows a "parabolic bell" shape.

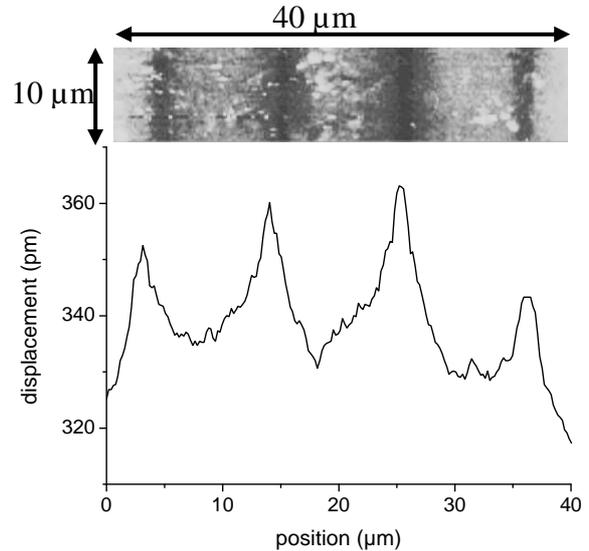

Fig. 3. SJEM expansion amplitude image and section (sample A).

From Fig. 3, we can also deduce that the measured resistor mean width, estimated by the full width half maximum, is 1.8±0.2 µm instead of the expected 350 nm. Several phenomena can explain it. First, when a device is functioning, the spatial limitation comes from the thermal one because even at 1 MHz, the thermal diffusion length is a few micrometers (5 µm for silicon). So the device shape is enlarged by the thermal phenomenon and if two thermal sources are, for instance, 1 µm away one from another, they will not be individually distinguishable. At f=10 kHz, this issue is even more critical as the thermal diffusion length is about 50 µm for silicon. In addition, this phenomenon is worsened by the fact that the tip is sweeping the passivation layer surface and not the active zone itself. So, the expansion signal due to the heating resistors is "filtered" and then smoothed by the passivation layer. The image consequently corresponds to an expansion mapping of the surface and not of the active zone itself. At last, when working under normal pressure conditions, the liquid meniscus effect between the tip and the sample surface is also a spatial resolution limiting effect as it constitutes the actual contact surface larger than the tip contact surface[10,16].

Another image has been obtained in the same experimental conditions apart for the excitation frequency f=1 kHz instead of 9 kHz. The measured resistor mean width, estimated by the full width half maximum, is then 2±0.2 µm. It is slightly higher than for 9 kHz, which can be explained by the higher thermal diffusion length or the measurement uncertainty.

Finally, we have analyzed, in Fig. 4, the expansion signal amplitude variations as a function of the power dissipated in one of the resistors for an excitation frequency f=10 kHz (same experimental conditions as in Fig. 3). We clearly see a linear behaviour, hence a quadratic dependence of the expansion amplitude on the supplying voltage amplitude $V_0$ as expected according to (2).

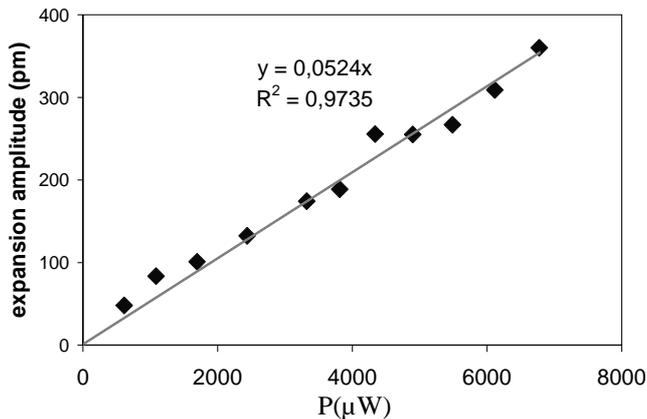

Fig. 4. SJEM expansion amplitude as a function of the power dissipated in each resistor (sample B).

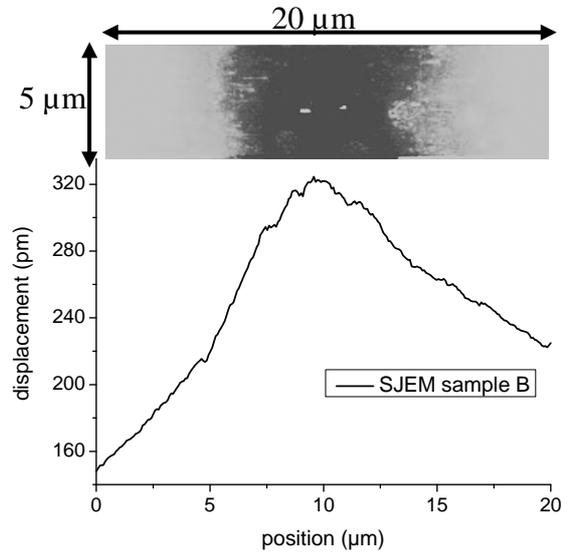

Fig. 5. SJEM expansion amplitude image and section (sample B).

*B. SJEM on Sample B*

With f=10 kHz and $V_0$=2.5 V, we have measured the signal on sample B on a 20 µm×5 µm area corresponding to 64 lines of 256 points each (dashed area in Fig. 2(b)). The SJEM image and the amplitude section are presented in Fig. 5.

The displacement is in pm. In this case, the thermal diffusion length is far bigger than the distance between 2 resistors (0.8 µm) and even bigger than the width of the whole structure made of the nine resistors (about 7 µm). Therefore, we only detect a global displacement of the whole structure and not the individual displacement of each resistor. The maximum displacement is 320 pm on the centre of the structure.

IV. COMPARISON WITH OPTICAL INTERFEROMETRY AND SCANNING THERMAL MICROSCOPY MEASUREMENTS

*A. Optical Interferometry Imaging*

We have also measured the thermal expansion using a classical Michelson interferometry technique[14] coupled to fast galvanometric mirrors to build the image (Fig. 6).

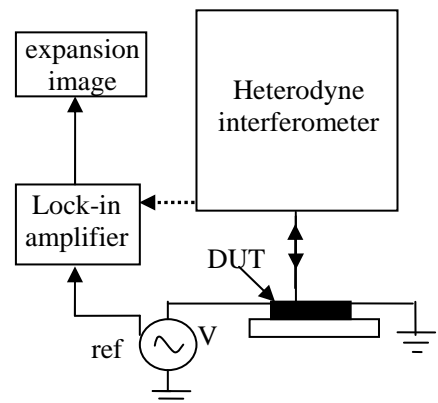

Fig. 6. Interferometer set-up

We have used a commercial heterodyne interferometric probe to measure the expansion due to the Joule effect: the reference and probe beams are modulated at two different frequencies and the detection is made at the frequency difference. Due to the acquisition system[17], this set-up is at present limited to displacement imaging for f frequency higher than 20 kHz. The laser probe after reflection on the sample interferes with a reference arm. The signal detected is sent into the lock-in amplifier locked on the f frequency.

*a. Interferometry on Sample A*

We present in Fig. 7 the image obtained on a 100 µm×100 µm area corresponding to 250 lines of 250 points each with $V_0$=2.5 V and f=50 kHz. We clearly detect the displacement on the resistors due to the Joule effect. Fig. 8 corresponds to the mean amplitude along sections between points A and B on the 250 lines. Even if the results are not rigorously comparable because of the thermal frequency difference (50 kHz instead of 10 kHz), we must note the same behaviour in Fig. 8 than in Fig. 3. Nevertheless, we must notice two main differences. The first one is the displacement amplitude on the resistors which is weaker than the one measured in Fig. 3. The second difference is the lower global displacement amplitude, here 40 pm instead of 320 pm in Fig. 3.

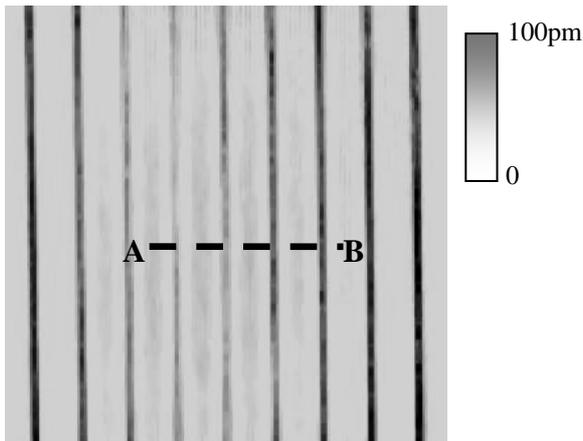

Fig. 7. Interferometric expansion amplitude image, f=50 kHz (sample A).

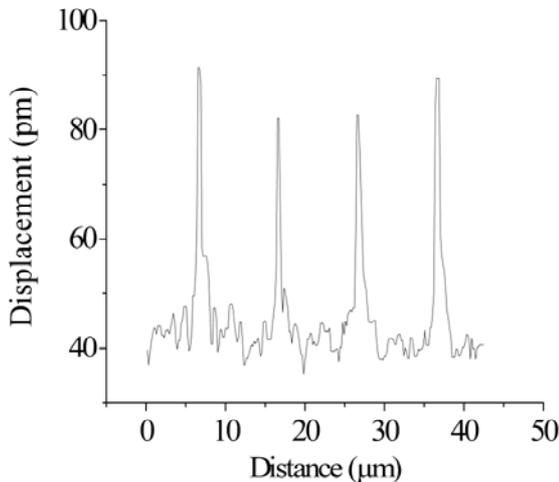

Fig. 8. Interferometric expansion amplitude section, f=50 kHz (sample A).

One explanation may be the higher excitation frequency which results in a weaker temperature variation amplitude and hence in a weaker surface displacement. It can also be explained by the fact that we do not exactly measure the same expansion. Using SJEM, the tip scans the surface of the sample whereas, using interferometry, the probe beam goes through the visible light transparent passivation layer and is directly reflected at the surface of the resistors. Consequently, in the first case, the method is sensitive to the expansion of the surface of the sample while, in the second case, the method measures the expansion of the surface of the resistors underneath.

*b. Sensitivity Comparison of SJEM and Interferometry on Sample B*

To compare both methods, we have measured, on the centre of sample B, the displacement created by a thermal excitation frequency f=10 kHz and with various amplitude values $V_0$. In this case, the laser probe is focused on one point, consequently we do not use the galvanometric mirrors but a classical point interferometric system[14] and hence we can work at frequencies lower than 20 kHz. We have superimposed in Fig. 9 the results obtained with SJEM with the ones obtained with interferometry. In both cases, we note comparable displacement values which, as expected, decrease linearly with the power for powers above 200 µW. Below this value, we see that the SJEM signal does not vary anymore while the interferometric signal goes on decreasing. SJEM images confirm that, below 200 µW, the signal is only noise. The sensitivity limitation of SJEM measurements is therefore a few picometers whereas it can reach less than 10 fm with interferometry [14].

Concerning the spatial resolution, Fig. 8 presents a shaper profile than Fig. 3 and we estimate the full width half maximum on the resistors at 600 nm. This value is far lower than the 2 µm measured with SJEM. This is explained by the method itself which measures directly the expansion on the resistors and by the higher frequency which reduces the thermal diffusion length.

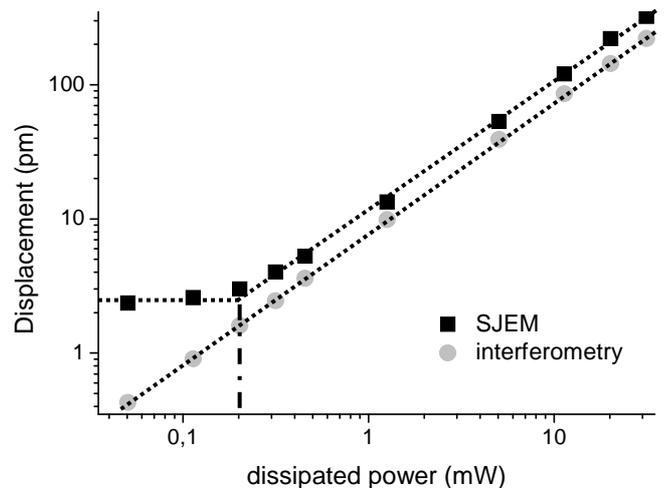

Fig. 9. displacement (pm) versus power (mW) for SJEM and interferometric measurements (sample B).

## B. Scanning Thermal Microscopy (SThM) Imaging

At last, we can compare the SJEM images obtained with the topographical probe and thermal images obtained in SThM with a Wollaston probe. This probe is constituted of a Wollaston wire shaped into a tip and etched to uncover its core platinum (Pt). The Pt wire is included in a Wheatstone bridge and used as a thermistor, as the resistance of the wire depends on its temperature. Such as in SJEM technique, the SThM is used in an AC regime. As a consequence, we use a lock-in scheme to measure the amplitude of the first harmonic in the tip voltage which is proportional to the tip mean temperature variations [12]. Consequently, from these measurements, we can deduce quantitative temperature variations of the tip and qualitative temperature variations at the surface of the sample. The bridge voltage is measured while scanning the surface so that the Pt electrical resistance could be estimated at each point of the sample surface. We estimate the diameter of the tip to be of the order of 5 µm and its contact radius of the order of 50 nm. Therefore, spatial resolution better than the one reached with thermoreflectance techniques can be expected[9]. We have already used this technique in AC regime[11] and we have in particular measured quantitative temperature variations on PN $Bi_2Te_3$ thermoelectric couples[12].

Here, the resistors are submitted to a 1 kHz sine 0 to 5 V positive voltage and consequently, the power dissipated by Joule effect at frequency f is 6.8 mW in each resistor. The scan size is 40 µm×10 µm, the scan rate 0.2 Hz and a line is constituted of 256 samples. The tip mean temperature variation image obtained is presented in the upper part of Fig. 10. Under the image, we present the signal measured on a section of this image. It actually presents the mean signal obtained on the 64 lines constituting the image.

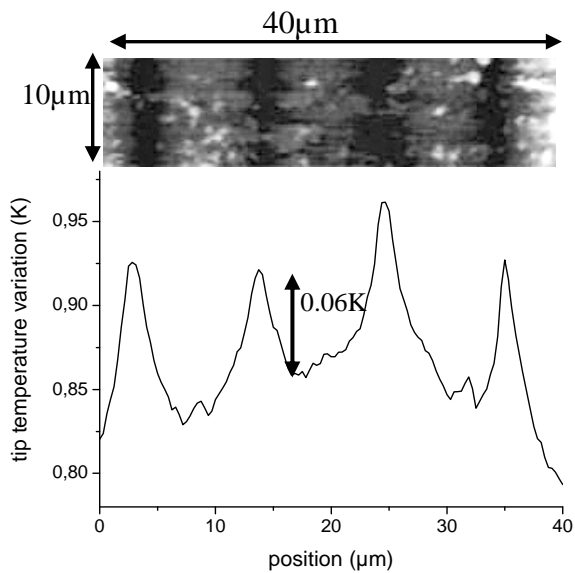

Fig. 10. SThM thermal image and mean temperature variation section signal, f=1 kHz (sample A).

In Fig. 10, we clearly see 4 heating resistors. The amplitude represents the mean tip temperature variations. The spacing between two resistors corresponds to the expected 10.35 µm. But the width of each heating resistor is 3±0.3 µm instead of the expected 350 nm. Again, as explained in section III for SJEM, the spatial limitation comes from several phenomena: the thermal diffusion length which is of the order of 150 µm for silicon at 1 kHz, the liquid meniscus effect between the tip and the sample surface and the measurement itself which is made on the surface of the passivation layer. The image consequently corresponds to a qualitative temperature variation mapping of the surface and not of the active zone[18]. But the value of 3 µm is also higher than the 2 µm one obtained with SJEM at the same f excitation frequency. We assume that it is also due to the geometry and size of the tip.

Apart for the tip geometry, the main drawback of SThM measurements in comparison with SJEM ones is the tip cut-off frequency estimated to 857±20 Hz[19] whereas the topographical tip cut-off frequency is several tens of kHz. Indeed, at 9 kHz, the thermal signal was too weak to obtain a satisfactory SThM image. As a consequence, since in thermal or expansion measurements, the spatial resolution is limited by the thermal diffusion length, it is in our interest to work at high frequency, thus to use SJEM instead of SThM, to improve the resolution. Another difficulty is to deduce quantitative temperature variations of the active zone as all the thermal exchanges between tip and surface must be known and evaluated and the influence of the passivation layer covering the active zone must also be estimated.

## C. Discussion: Performances of SJEM, Interferometry and SThM Measurements

To sum up, we present in table 1 the performances of the three methods in terms of spatial resolution, sensitivity and bandwidth for the samples studied and under our experimental conditions.

Concerning the spatial resolution for SJEM and SThM methods, even if the contact radius of the tip is clearly submicrometric (50 nm), the limitation is a thermal one because of the thermal diffusion length. The spatial resolution is better with interferometry because of the higher thermal excitation frequency and because this method measures the displacement on the resistors themselves and not on the covering passivation layer.

TABLE I
PERFORMANCES OF SJEM, INTERFEROMETRY AND STHM

| Technique | SJEM | interferometry | SThM |
|---|---|---|---|
| Spatial resolution | 2µm | 600 nm | 3µm |
| Sensitivity | <10pm | <10 fm | <0.1 K |
| Frequency range | >10kHz (deflection) >200Hz and <10kHz (topo) | <1MHz >20kHz(scan) | >200Hz <1 kHz |

According to Fig. 10, the SThM sensitivity is better than 0.06 K on this sample but we have measured lower temperature variations on other samples. As for surface displacements, the sensitivity is a few picometers for SJEM and a few femtometers for point interferometry.

Concerning the frequency bandwidth, it is highly dependent on the experimental conditions. For SJEM using the deflection image, we must use frequencies higher than the bandwidth of the feedback loop of the AFM, here 10 kHz. But experimentally, we know that the thermal signal decreases when the frequency increases. So, in practice we are limited, for our samples, to frequencies lower than a few tens of kHz. We must also be aware of resonance frequencies of the tips. We can work at frequencies lower than 10 kHz using the topographical image. But for all the AFM methods using modulated excitation and a lock-in amplifier, here SJEM and SThM, the low limit is a few hundreds Hz because of the scan rate of the AFM whose lower limit is 0.2 Hz per line, thus 50 Hz per point (with 256 points per line). As for SThM, the high frequency limitation (1 kHz) comes from the tip cut-off frequency. We can do measurements at higher frequencies but the signal will be attenuated. At last, concerning interferometry, if we use point measurements, the only limitation is a high frequency one depending on the photodetector bandwidth, typically 1 MHz for the sensitivity given in [14]. But with a scanning system, in addition to the 1 MHz high frequency limitation, there is a low limitation due to the scan speed of the galvanometric mirrors and to the use of a lock-in detection, which imposes to work above 20 kHz.

## V. CONCLUSION

We have presented surface displacement images obtained on two electrical devices using two different techniques: an optical interferometric one and a Scanning Joule Expansion Microscopy one.

In addition, we have presented SThM images to compare them with SJEM images. We have underlined that the main drawback of SThM measurements regarding SJEM ones is the geometry and the low cut-off frequency of the tip that demands low frequency measurements and limits the spatial resolution. We have presented the performances of the three methods in terms of bandwidth, spatial resolution and sensitivity on the same device.

We are now studying the influence of the working frequency on the spatial resolution and we plan to work under vacuum to improve it.

ACKNOWLEDGMENT

The samples have been provided by the GDR Micro- et nanothermique (n°G2503).

This work has been supported by the ANR PNANO "EXTRADA".